\documentclass{article}
\pdfoutput=1

\PassOptionsToPackage{numbers, compress}{natbib}

\usepackage[preprint]{nips_2018}

\usepackage[utf8]{inputenc} 
\usepackage[T1]{fontenc}    
\usepackage{hyperref}       
\usepackage{url}            
\usepackage{booktabs}       
\usepackage{amsfonts}       
\usepackage{nicefrac}       
\usepackage{microtype}      
\usepackage[disable]{todonotes}
\usepackage{comment} 
\usepackage{multirow}
\usepackage{amsthm}
\newtheorem{definition}{Definition}
\newtheorem{theorem}{Theorem}
\newtheorem{lemma}{Lemma}

\usepackage{subcaption}
\usepackage{graphicx}
\usepackage{amsmath,amssymb,latexsym,exscale}
\usepackage{bbold} 
\usepackage[american]{babel}

\usepackage{booktabs}
\usepackage{tabularx}

\usepackage{capt-of}

\usepackage{algorithm}
\usepackage{algpseudocode}
\algnewcommand\algorithmicinput{\textbf{Input:}}
\algnewcommand\Input{\item[\algorithmicinput]}
\algnewcommand\algorithmicoutput{\textbf{Output:}}
\algnewcommand\Output{\item[\algorithmicoutput]}

\usepackage{xspace}
\newcommand{\Lall}{\mathcal{L}}
\newcommand{\Lallt}{\Lall_{t'}}

\newcommand{\iid}{i.\,i.\,d.\xspace}

\newcommand{\eps}{\epsilon}
\newcommand{\wrt}{w.\,r.\,t.\xspace}

\graphicspath{{figures/}}
\DeclareGraphicsExtensions{.pdf}
\newcommand{\figurewidth}{.9\columnwidth}

\title{The Influence of Differential Privacy on \\Short Term Electric Load Forecasting}

\author{
  G{\"u}nther~Eibl%
  \\
  Salzburg University of Applied Sciences\\
  Center for Secure Energy Informatics\\
  \texttt{guenther.eibl@en-trust.at} \\
  \And
  Kaibin~Bao \\
  Karlsruhe Institute of Technology (KIT) \\
  Institute AIFB \\
  \texttt{kaibin.bao@kit.edu} \\
  \And
  Philip-William~Grassal\\
  SAP Security Research\\
  \texttt{philip-william.grassal@sap.com} \\
  \And
  Daniel~Bernau\\
  SAP Security Research\\
  \texttt{daniel.bernau@sap.com} \\
  \And
  Hartmut~Schmeck\\
  Karlsruhe Institute of Technology (KIT) \\
  Institute AIFB \\
  \texttt{hartmut.schmeck@kit.edu} \\
}

\begin{document}

\maketitle

\begin{abstract} 
	There has been a large number of contributions on privacy-preserving smart metering with Differential Privacy, addressing questions from actual enforcement at the smart meter to billing at the energy provider.
	However, exploitation is mostly limited to application of cryptographic security means between smart meters and energy providers.
	We illustrate along the use case of privacy preserving load forecasting that Differential Privacy is indeed a valuable addition 
	that unlocks novel information flows for optimization.
	We show that (i) there are large differences in utility along three selected forecasting methods, (ii)  energy providers can enjoy good utility especially under the linear regression benchmark model, and (iii) households can participate in privacy preserving load forecasting with an individual re-identification risk $< 60\%$, only $10\%$ over random guessing.
	This is a pre-print of an article submitted to Springer Open Journal ``Energy Informatics''.
\end{abstract}


\section{Introduction}

Smart metering data is said to be useful for improving the load forecasting task of energy providers \cite{McDaniel09,Li10a,Ilic2013_impact,Bao15a}.
With more accurate forecasts, energy providers gain an advantage for trading and scheduling electricity production and consumption ahead of time.
Forecasting errors have to be balanced with control energy for stable electric grid operation.
Thereby, the highly volatile control energy prices charged for this compensation can be painful for the energy providers.
In Germany of 2017, for example, the average control energy price was 49.67 EUR per MWh, but for 30 minutes, the price shot over 20,614.97 EUR per MWh.\footnote{The control energy price in Germany (``reBAP'') is available for download on \url{www.regelleistung.net}}

On the other hand, monitoring electrical load from individual households incurs violation of privacy,
as private behavior patterns are reflected in the energy consumption\footnote{e.g., household occupancy, appliance usage or approximate sleep-wake-cycles} \cite{McDaniel09,Molina10,Lisovich2010inferring}. 
The amount of privacy violation varies depending on the monitoring time  resolution of metering data \cite{Eibl15a}.
Using Differential Privacy \cite{Dwork06b} as privacy model, both time granularity and varying levels of the privacy parameters can be used to quantify and interpret the influence on privacy.

In addition to the privacy issue, the utility of individual data for load forecasting is naturally limited due to the stochasticity of domestic energy usage \cite{Fan09multiregion}.
To best of our knowledge, no work exists that 
leverages individual (instead of aggregated) load data to gain a significant advantage on the domestic load forecasting task.
This is why domestic load forecasting is performed using load data aggregated over large areas with many households.

In this paper, we investigate whether energy providers can benefit from smart metering data which is acquired in a privacy-friendly way.
We formulate a privacy-preserving forecasting process that provides energy producers with forecast utility guarantees and households with strong, yet intuitive privacy guarantees, based on Differential Privacy.
We make the following contributions:
\begin{itemize}
	\item First time to regard energy provider's load forecasting task based on smart metering data with prescribed privacy guarantee,
	\item Practical design and evaluation of Differential Privacy for load forecasting, as well as
	comprehensible and interpretable calculation of presence detection risk using Differential Identifiability,
	\item Determination of the privacy-utility trade-off on real world data \cite{Hong12} using three 
	realistic 
	forecasting methods, and
	\item Demonstrating that differentially private load forecasting with a low presence detection risk $\rho < 0.6$ and strong utility is especially achievable under the linear regression benchmark model.
\end{itemize}

This paper is structured as follows.
In Section~\ref{sec:prelim}, we introduce preliminaries.
We formulate our concept for realizing differentially private load forecasting in Section~\ref{sec:problem} and
present an evaluation in Section~\ref{sec:experiments}.
Related work is presented in Section~\ref{sec:related}.
Finally, Section~\ref{sec:Conclusion} concludes with a discussion of practical implications.


\section{Preliminaries}
\label{sec:prelim}
In the following, we provide fundamentals regarding electricity grid metering (Section~\ref{subsec:emetering}), the underlying privacy model of this work (Section~\ref{subsec:differential-privacy}), and load forecasting approaches we use for electricity consumption prediction (Section~\ref{sec:ForecastMethods}).

\subsection{Electricity Metering Process (in Germany)}
\label{subsec:emetering}

In this paper, we will discuss our problem setting in the context of the German metering and balancing process.
Although the objective of metering for balancing in an electric power system is equal around the world,
specific details in metering and settlement are subject to national and regional regulations.
That is why we fix our process description to the well-documented German electrical power market.
The relevant sources are the German electricity grid access regulation (StromNZV~\cite{stromnzv}),
the German measuring point operation act (MsbG~\cite{msbg})
and the market rules for the implementation of balancing group accounting for electricity (MaBiS~\cite{mabis}).

In Europe, the electric grid is partitioned geographically into control areas which are each operated by a transmission system operator (TSO).
Each control area is subdivided into distribution grids operated by a distribution system operator (DSO).
Transmission and distribution grid operators are government-regulated entities who are responsible for stable and reliable grid operation and non-discriminatory access to electricity production, consumption and trading.
To accomplish these two goals simultaneously, the TSO delegates the task of balancing supply and demand to the grid participants to some extent by charging the participants for any imbalance they cause.
How imbalance is estimated and settled is subject to national regulations.
(cf. \cite{ec_2017_1485,ec_2017_2195,federal2015energy})

\begin{figure}[htb]
	\centering
	\includegraphics[width=\figurewidth]{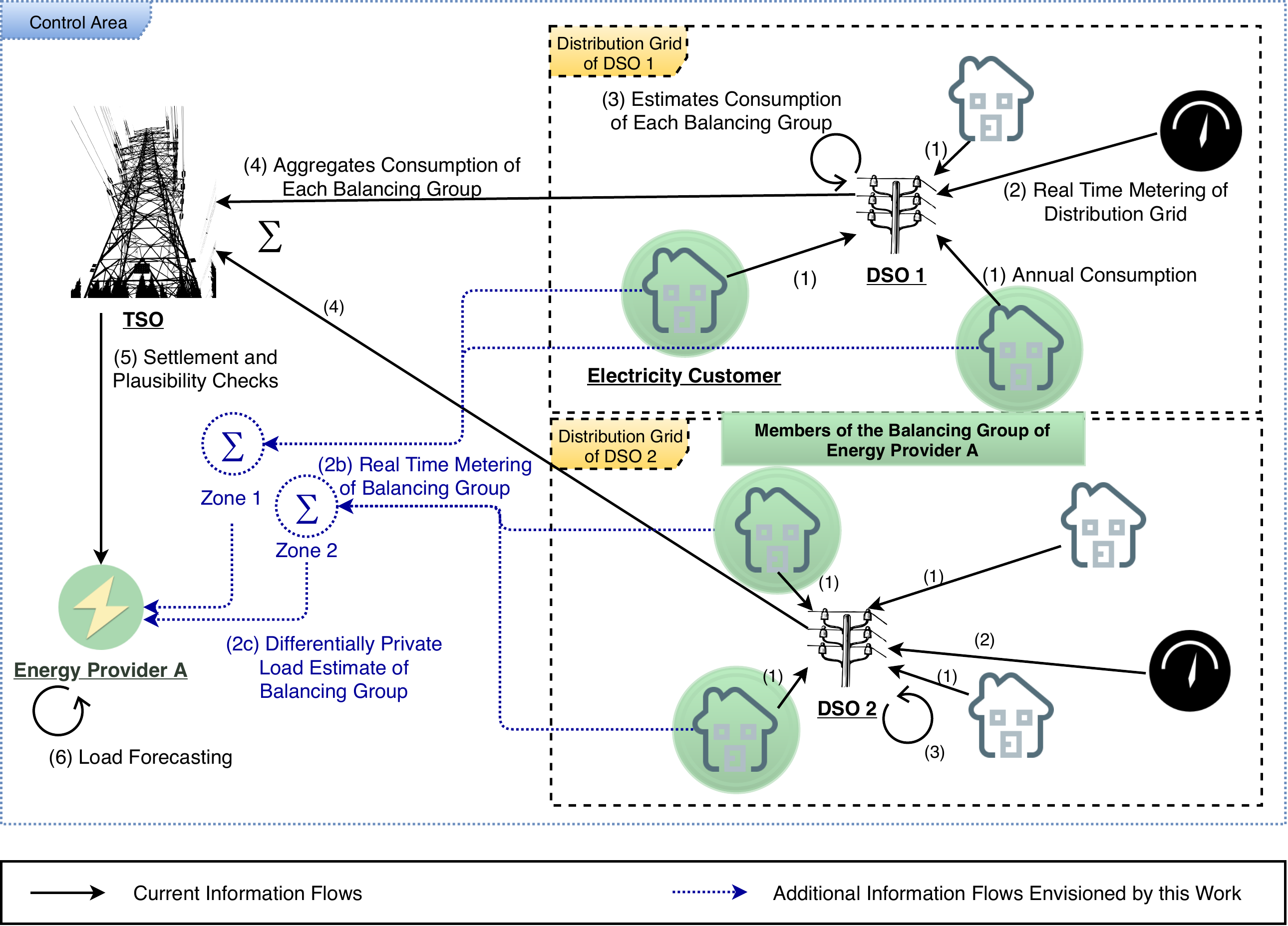}
	\caption{Essential roles and information flows in the current metering process (solid black arrows) and in the proposed differential private process for privacy-preserving improved forecasting (dotted blue arrows). For simplicity, we omitted the role of the Measurement Service Provider who is currently responsible for step 1.}
	\label{fig:roles_and_info_flows}
\end{figure}

Each control area is virtually partitioned into balancing groups which are basically time-dependent accounts for electric energy.
An electricity customer (i.e., her grid connection point) is associated with exactly one balancing group which corresponds to the energy service provider and possibly to a specific tariff chosen by the customer.
(cf. Sections 4, 5 StromNZV)

Before the roll-out of smart metering, residential electricity meters of customers with low or normal annual consumption have only been read-out annually or during the change of energy provider or tenant.
Customers with an annual consumption above 100,000~kWh are subject to real-time load profile measurements which collects average and peak load in each quarter-hour interval.
With the roll-out of smart metering, additionally, customers with an annual consumption between 10,000~kWh and 100,000~kWh may be subject to load profile metering with quarter-hourly resolution.
(cf. Sections 55, 60 MsbG)

Figure~\ref{fig:roles_and_info_flows} shows the essential roles and information flows, as well as our envisioned privacy-preserving information flow, in the metering and balance settlement process.
The TSO is usually also in the role of the balancing group coordinator and is responsible for determining the virtual balance of each balancing group in order to charge the balancing group which is the responsible party for imbalances.
As the balancing group may be physically scattered among different distribution grids, the TSO needs to aggregate the information about the energy flows in the distribution grids from several DSOs.
The problem here is that the DSO does not measure every grid connection point in real-time.
This is especially true for residential grid connection points.
Therefore, the DSO estimates the residential loads either by using the synthetic or the analytical method (Step 1 and 2 in Figure~\ref{fig:roles_and_info_flows}).
The synthetic method uses parameterized standard load profiles which are scaled by a forecasted annual energy consumption of each customer.
For the analytical method, the DSO subtracts the real-time metered load profiles and estimated transmission losses from the overall load profile of its distribution grid.
The remainder is the load profile of the non-metered residential grid connection points, which is then attributed according to a forecasted annual energy consumption of each customer (Step 3 in Figure~\ref{fig:roles_and_info_flows}).
(cf. Section 1.2 MaBiS~\cite{mabis} and Section 3.8 of the Distribution Code 2007~\cite{distributioncode2007})

The TSO finally aggregates the load profiles from all DSOs to determine the load profile of each balancing group (Step 4 in Figure~\ref{fig:roles_and_info_flows}).
This overall ex-post balance in each group is used to settle the costs for the actual imbalance during the grid operation.
If the imbalance of one group helps to compensate the overall grid imbalance, the responsible party of the group is being paid for the grid support.
The parties responsible for the balancing group receive the load and balance measurements for their balancing group in order to retrace the bill and to improve on the load predictions for ex-ante energy trading (Step 5 in Figure~\ref{fig:roles_and_info_flows}).
(cf. Section 2 MaBiS)

Technically, the current metering process is not differentially private as the aggregated load of a balancing group is not perturbed using a randomized method.
Even the collection of the annual energy consumption is not differentially private.
However, the current metering process based on non-smart meters is generally not considered as serious privacy violation since residential electricity measurements are read out only once per year.

\subsection{Differential Privacy}
\label{subsec:differential-privacy}

Differential Privacy, originally proposed by Dwork \cite{Dwork06b}, is the current gold standard for data privacy.
It is achieved by perturbing the result of a query function $f(\cdot)$
s.t. it is no longer possible to confidently predict whether the result was obtained by querying data set $D_1$ or some other data set $D_2$  differing in one individual. Thus, privacy is provided to each participant in the data set as their presence or absence becomes almost negligible for computing perturbed query results. To inject noise into the result of some arbitrary query $f(\cdot)$, \textit{mechanisms} $K_f$ are utilized. Mechanisms add noise sampled from a probability distribution to $f(\cdot)$ and are differentially private when they fulfill Definition~\ref{def:differential-privacy}.
\begin{definition}[Differential Privacy]
	\label{def:differential-privacy}
	A mechanism \\$K_f:DOM \to R$
	is ($\epsilon$, $\delta$)-differential\-ly private if for all data sets $D_1,D_2 \subset DOM$ differing in only one individual and for all possible outputs $S \subseteq R$ :
	\begin{equation} \label{eq:DP}
	\centering
	Pr[K_f(D_1) \in S] \le e^\epsilon * Pr[K_f(D_2) \in S] + \delta \quad .
	\end{equation}
\end{definition}
The additive $\delta$ is interpreted as the probability of protection failure and required to be negligibly small $\approx \frac{1}{|D_1|}$. We refer to Dwork et al. \cite{Dwork13a} for the proof. Another commonly used, more strict definition calls a mechanism $\epsilon$-differential\-ly private if it is ($\epsilon, 0$)-differential\-ly private. Differential Privacy has the appealing property that 
	it holds independent of side knowledge that an adversary might have gathered on the data set.
Thus, for convenience, we call a data set differentially private if it has been obtained by a differentially private mechanism. 

The query is further specified as a series of $k$ identical aggregate queries $f_i$ with co-domain $R=\mathbb{R}$ each. The added noise must hide  the influence of any individual in the original result of the composed query  $f=(f_1,\ldots,f_k)$. The maximum influence of an individual on 
$f(\cdot)$ is the \textit{global sensitivity} $\Delta f = \max_{D_1,D_2} \|f(D_1)-f(D_2)\|_1$.

A popular mechanism for perturbing the outcome of numerical query functions is the \textit{Laplace mechanism}, proposed by Dwork \cite{Dwork06a}. It adds noise calibrated \wrt the global sensitivity by drawing a random sample from the Laplace distribution with mean $\mu = 0$, scale $\lambda = \frac{\Delta f}{\epsilon}$ 
according to Theorem~\ref{def:lap-mech}.
\begin{theorem}[Laplace Mechanism]
	\label{def:lap-mech}
	Given a series of $k$ identical numerical query functions $f=(f_1,\ldots,f_k)\in\mathbb{R}^k$, the Laplace Mechanism 
	\begin{equation} \label{eq:Laplace}
	K_{Lap}(D,f,\epsilon) := f(D) + (z_1, ..., z_k) 
	\end{equation}
	is an ($\epsilon$,0)-differentially private mechanism	
	when all $z_i$ with $1 \le i \le k$ are independently drawn from the random variable $\mathcal{Z} \sim Lap(z,\frac{\Delta f}{\epsilon},0)$.  
\end{theorem}

Again, for proof, we refer to Dwork et al. \cite{Dwork06a}.
To apply Theorem~\ref{def:lap-mech} to smart metering, i.e., a distributed setting, we use the gamma distribution suggested for distributed noise generation by \'Acs et al. \cite{Acs11a}.
The following Lemma~\ref{def:gamma} leads to the generation of gamma noise that satisfies the Laplace mechanism.
We use this divisibility to formulate a distributed differentially private metering process in Section~\ref{sec:differentially-private-metering-process}.

\begin{lemma}[Divisibility of Laplace distribution \cite{Kotz2001,Acs11a}]  \label{def:gamma}
Let $\mathcal{Z}(\lambda)$ denote a random variable from a Laplace distribution with density $f(x, \lambda) = \frac{1}{2\lambda}e^{\frac{|x|}{\lambda}}$. 
Then the distribution of $\mathcal{Z}(\lambda)$ is 
infinitely divisible. This means that for every integer $n \ge 1$ it can be represented as a sum of $n$ random variables $\mathcal{Z}(\lambda) = \sum_{i=1}^{n}X_i$. Here, each $X_i=\mathcal{G}_1(n, \lambda) - \mathcal{G}_2(n, \lambda)$. $\mathcal{G}_1(n,\lambda)$ and $\mathcal{G}_2(n,\lambda)$
are \iid random variables having gamma distribution with density $g(x, n, \lambda) = \frac{(1/\lambda)^{1/n}}{\Gamma(1/n)}x^{\frac{1}{n}-1}e^{-x/\lambda}$ defined for $x \ge 0$.
\end{lemma}

When a function is evaluated multiple times 
an overall \textit{privacy loss} occurs.  
Under worst case assumptions, the \textit{sequential composition theorem} of Differential Privacy states that a series of $k$ evaluations of any ($\epsilon,~\delta$)-differentially private mechanism $K_f$ on the same set of individuals results in ($k\epsilon$,~$k\delta$)-Differential Privacy. However, recent results by Dwork et al. \cite{Dwork10a} and Kairouz et al. \cite{Kairouz2017} prove that actually sub-linear increases in $\epsilon$ are achieved under $k$-fold composition when allowing a small $\tilde{\delta}$ under Theorem~\ref{def:adv-composition}.
\begin{theorem}[$k$-Fold Adaptive Composition for Homogeneous Mechanisms]
\label{def:adv-composition}
	For any $\epsilon > 0$ and $\delta \in [0,1]$, and $\tilde{\delta} \in (0,1]$ the class of ($\epsilon$, $\delta$)-differentially private mechanisms satisfies ($\tilde{\epsilon}_{\tilde{\delta}}$, $1-(1-\delta)^k(1-\tilde{\delta})$)-Differential Privacy under k-fold adaptive composition, for 
\begin{align} \label{eq: epscomposition}
\tilde{\epsilon}_{\tilde{\delta}} = \min \begin{cases}
	k \epsilon & \\ 
	\frac{(e^{\epsilon}-1)k\epsilon}{e^{\epsilon}+1} + \epsilon\sqrt{2k\ln\left(e+\frac{\sqrt{k\epsilon^2}}{\tilde{\delta}}\right)} &\\
	\frac{(e^{\epsilon}-1)k\epsilon}{e^{\epsilon}+1} + \epsilon\sqrt{2k\ln\left(\frac{1}{\tilde{\delta}}\right)}
\end{cases} \quad .
\end{align}
	
\end{theorem}
When operating in high privacy regimes ($\epsilon \ll 1$), the term $\frac{(e^{\epsilon}-1)k\epsilon}{e^{\epsilon}+1} \approx k\epsilon^2$ illustrates the sub-linear loss of privacy under $k$-fold composition.
Even though composition allows to determine the privacy decay by growth in $\epsilon$ over a series of queries, a rational explanation for the actual choice of $\epsilon$ is missing.
To the best of our knowledge, there is no approach for giving concrete guidance on choosing $\epsilon$. Nonetheless, we are convinced that providing a more comprehensible interpretation of $\epsilon$ and the corresponding guarantee is crucial for  acceptance of Differential Privacy in practice. 

Consequently, we apply a \textit{belief model} in this work to give smart metering users a better understanding of their protection guarantee $\epsilon$.
The foundation of this model led Lee et al. \cite{Lee12} to define \textit{Differential Identifiability}, a privacy notion slightly differing from \textit{Differential Privacy}. For convenience, we restate the definition of Differential Identifiability in Definition~\ref{def:diff-identifiability}.
 
\begin{definition}[Differential Identifiability]
\label{def:diff-identifiability}
	Given an original data set $D$, a randomized mechanism $K$ satisfies $\rho$-Differential Identifiability if among all possible databases $D_1, D_2, ..., D_m$ differing in one individual \wrt $D$ the posterior belief after getting the response $r$ 
	\begin{equation}
	P(D_i | K(D) = r) \le \rho
	\end{equation}
	is bounded by $\rho$.
\end{definition}
$\rho$-Differential Identifiability implies that after receiving a mechanism's output $r$ the true data set $D$ can be identified by an adversary with confidence $\le \rho$.
Findings by Li et al. \cite{Li13} show that Differential Privacy and Differential Identifiability are actually equal when $m = 2$ since Differential Privacy considers only two neighboring data sets $D_1$, $D_2$ by definition. If this condition is met, according to Li et al. \cite{Li13}, the relation between $\rho$ and $\epsilon$ is:
\begin{align}
\label{equ:eps-rho}
	\eps = \ln\left(\frac{\rho}{1-\rho}\right) \quad \text{and} \quad \rho = \frac{1}{1+e^{-\epsilon}} > \frac{1}{2}
	\quad .
\end{align}
Consequently, the re-identification confidence $\rho$ provides a simplified interpretation of the actual risk when applying $(\epsilon, 0)$-Differential Privacy. When $\delta > 0$, we define that the confidence of $\rho$ holds with probability $1-\delta$.
We use this method to substantiate our results in Section~\ref{subsec:application-of-differential-private-mechanism}.

\subsection{Electric Load Forecasting Methods} \label{sec:ForecastMethods}

Three different forecasting methods are used within this work. One of the methods is the benchmarking forecasting model (Section~\ref{sec:modBenchmark}) for the 2012 Global Energy Forecast Competition (GEFCom 2012). The other two methods, \textit{CountingLab} (Section~\ref{sec:modelClab}) and \textit{Lloyd} (Section~\ref{sec:modelLloyd}), were the two highest ranked forecasting methods of the competition. 
For the first time, the impact of Laplacian noise for differential privacy on 
realistic forecasting methods is studied in Section \ref{sec:experiments} .

\subsubsection{Global Energy Forecast Competition 2012 (GEFCom 2012)} \label{sec:GefCom}

One of the machine learning competitions of GEFCom 2012 \cite{Hong12} was electric load forecasting.
The time span of the given historical load data of an ISO in the USA was approximately 4.5 years in hourly readout intervals 
from 20 zones.
Additionally, historical temperature data of 11 nearby weather stations were given, but there was no information about the association between weather stations and zones.
For the forecasting time period, the temperature data was not given and needed to be forecasted, too. A limited amount of tuning is possible due to allowing multiple submissions and directly showing the resulting score.

\begin{figure}[htb]
	\includegraphics[width=\figurewidth]{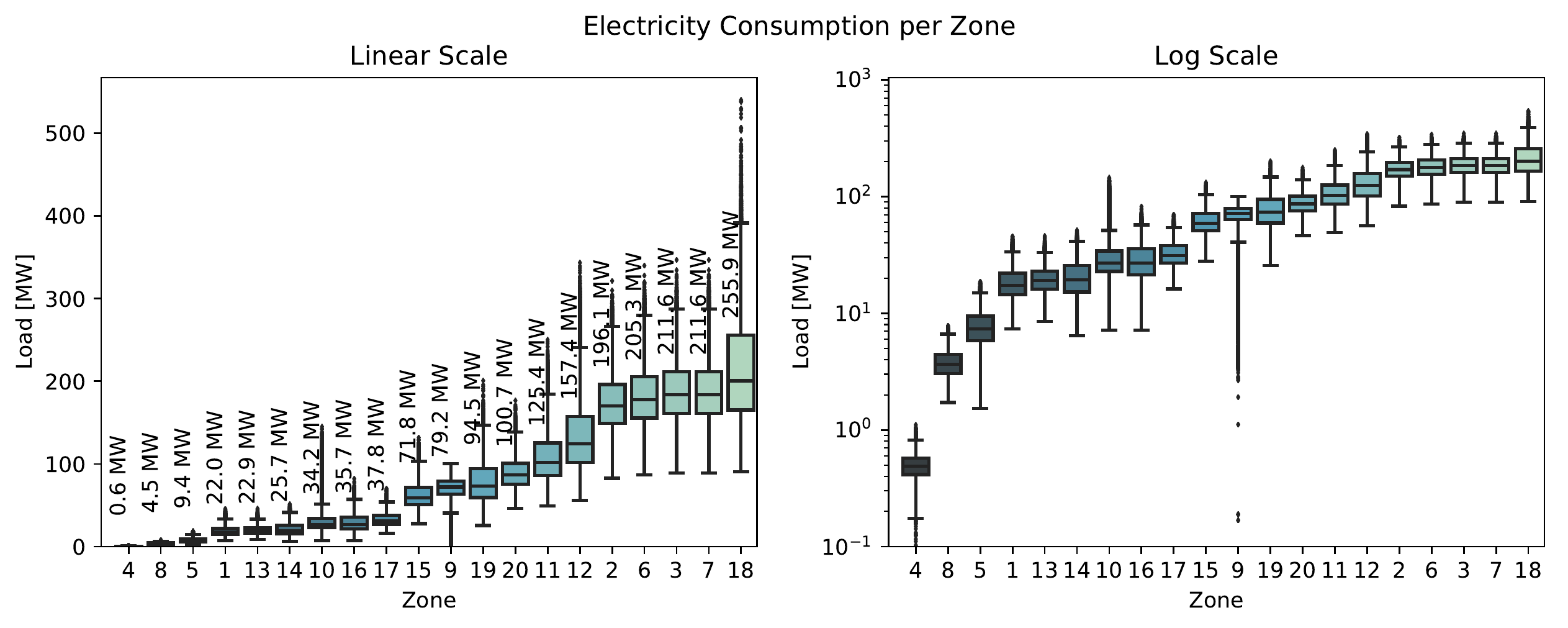}
	\caption{Statistics of the 20 zones of the GEFCom 2012 data set.}
	\label{fig:zonestats}
\end{figure}

The statistics of the historical load data are plotted in Figure~\ref{fig:zonestats}.
Zone 4 is the smallest zone with a mean load of only 0.575 MW.
In the right panel it can be seen that Zone 9 exhibits outliers with low consumption values which indicates metering issues or local blackouts.

\subsubsection{Benchmark Forecasting Model of GEFCom 2012} \label{sec:modBenchmark}

Hong \cite{Hong12} provided a linear regression model as a benchmark for GEFCom 2012 competition.
A linear regression model for load forecasting has the general form 
\begin{equation}
\label{eqn:mlr}%
F_{z,t} = \beta_0 + \sum_{j=1}^{p} \beta_j  x_{t,j} + e_t \quad ,
\end{equation}
where $F_{z,t}$ is the forecast of the aggregate energy consumption of zone $z$ in time slot $t$,
$\beta_j$ are the parameters of the model,
$x_{t,j}$ are the independent variables,
$e_t$ is the residual error which cannot be explained by the model.

The 20 benchmark models (one per zone) consider a total of $p = 313$ explanatory temperature and calendar variables $x_{t,j}$ or cross-effects which are described in the GEFCom 2012 paper \cite{Hong12}.

For each zone, only one of the 11 weather stations is chosen to provide the temperature values for the linear model.
The choice was made by fitting the model for each of the weather stations and choosing the one with the smallest training error $\sum_{t=1}^k |e(t)|$ where the errors are summed up for all time points.

Using that final model, a forecast for the week following the given historical data was to be estimated. While the explanatory calendar variables can be easily obtained, no  forecast for temperature $T_s$ of the weather stations was given.
The benchmark model constructed temperature forecasts by ``averaging the temperature at the same date and hour over the past four years'' \cite{Hong12}.

\subsubsection{CountingLab's Forecasting Method} \label{sec:modelClab}

Within the forecasting methods referenced in the GEFCom 2012 paper \cite{Hong12}, CountingLab \cite{Charlton14} achieved the best test score in the competition.
 As the benchmark model, it relies on multiple linear regression (\ref{eqn:mlr}). However,
in contrast 
to the benchmark model, not 20 single forecasts are obtained (one per zone) but 3,840 forecasts $F_{z,h,S,w}$ for each combination of zone $z$, hour of the day $h$, season $S$ and day type. 
As a benefit the number of independent variables is much smaller than for the benchmark model: only nine parameters (interactions of temperature, day number and day number within the season) must be fit per linear model. 

The needed temperature forecast is the mean of historical temperatures. 
In order to win the competition the authors spent additional effort, see Charlton et al. \cite{Charlton14} for more details.

\subsubsection{Lloyd's Forecasting Method} \label{sec:modelLloyd}
\label{sec:modelLloyd}

Lloyd's method \cite{Lloyd14} achieved the second best test score in the competition. 
First, the temperature was estimated as the sum of a smooth trend and a daily periodic estimate using Gaussian processes with squared exponential and periodic kernels, respectively.

The prediction is a weighted ensemble of three forecasting methods: (i) the benchmark model (see Section \ref{sec:modBenchmark}) with weight 0.1, (ii) a gradient boosting machine \cite{Friedman01a} with weight 0.765 and (iii) a Gaussian process regression with weight 0.135. The weights have been chosen by manual tuning.

For each zone, a separate boosting model was learned using as input the time of day $\in [0,1]$, the time within the week $\in [0,7]$, the temperature predictions and smoothed temperature predictions of all weather stations. Note that the loads are not used as inputs, only as response values.


The third method uses Gaussian process regression using three additive kernels for forecasting that all depend on time: two squared exponential kernels should explain the variation of the load by two different length scales; the third, periodic kernel should model the periodic behavior.


\section{Differentially Private Metering and Load Forecasting}
\label{sec:problem}

As we have discussed in Section \ref{subsec:emetering}, the energy provider has the incentive to forecast the load of her customers with low error so that she can trade or schedule energy production ahead of time for lower costs.
Smart metering may provide the means to realize a more accurate forecast by using load monitoring of individual households.
However, monitoring individual loads conflicts with customer privacy interests.
Also, due to the high stochasticity of individual loads, proficiently grouping households based on geographic or topological areas is beneficial to the forecasting performance (cf. Fan et al. \cite{Fan09multiregion}).

We propose that the energy provider and the customers meet in the middle by agreeing on a trade-off between forecasting accuracy and customer privacy.
For that, Differential Privacy is guaranteed for the customer by grouping households into zones and apply the Laplace mechanism on the aggregated load of each zone. In this paper we assume that a privacy-preserving protocol based on additional homomorphic encryption (\cite{Li10a, Erkin12a}) or masking (\cite{Acs11a, Knirsch16a}) exists in the smart metering infrastructure that enables to calculate the sum of all the household's load values at each time point without providing the individual values.

We envision that the energy provider offers different energy tariffs coupled with a specific privacy protection level in terms of different $\lambda$ values for the Laplace mechanism. 
The differentially private metering process is detailed in Section \ref{sec:differentially-private-metering-process}.

The customer interprets the privacy level coupled with the tariff by deriving her own effective privacy protection in terms of Differential Identifiability.
Using that assessment, she can perform an informed decision about her energy tariff and how much privacy she wants to trade-in.
This in described in Section \ref{subsec:application-of-differential-private-mechanism}.

From the perspective of the energy provider, the utility of individual smart metering data is limited. Since the load of the whole balancing group used for balance compensation must be provided by the TSO this could be used for a direct forecast of the aggregate of the balancing group. However, forecasting could possibly be improved by also providing differentially private, zonal sub-aggregates of the region obtained by the DSOs.
As shown in Section \ref{sec:experiments} depending on the forecasting method, using this so-called hierarchical forecast compared to the direct forecast may not even improve the forecasting performance.

If there is an advantage using the hierarchical forecast, the acquisition of smart metering data has to satisfy the privacy interests of the affected customers.
For each forecasting method, the limit for the privacy level of the differentially private method depends on the additional error introduced by it. If the forecasting error exceeds the direct forecasting error, smart metering data do not help the energy provider for trading or scheduling tasks.
In Sections \ref{subsec:forecasting_problem} and \ref{subsec:evaluation}, we reflect that idea in the definition of the load forecasting problem and the utility definition.

\subsection{Basic Forecasting Problem}
\label{subsec:forecasting_problem}

Consider a control area, called region for simplicity, that is divided into $Z$ zones containing $n_z$ households. All households $i$ of a zone $z$ provide their load measurements $l_{z,i,t'}$ at several time points $t'$. The zone aggregators (DSOs) calculate the sum of all the household's load values $L_{t'}$ at each time point $t'$ without receiving the individual values. 
Therefore, for each zone for each time point $t'$, only the sum of the load values $l_{z,i,t'}$ of all the households $i$ of the zone is available,
\begin{equation}
\label{equ:sum-zone}
L_{z,t'}:=\sum \limits_{i=1}^{n_z}l_{z,i,t'} \quad .
\end{equation}
These zonal aggregate loads are available at past time points $t'_1,\ldots,t'_k$. 
The goal then is to predict the regional aggregate load which is the sum of the zone's loads, 
\begin{equation}
\label{equ:sum-all}
\Lallt:=\sum \limits_{z=1}^{Z}L_{z,t'} \quad .
\end{equation}
Based on values available at times $t'_1,\ldots,t'_k$ the forecasting problem consists of producing forecasts $F_{t_1},\ldots, F_{t_T}$ for the regional aggregate load $\mathcal{L}$ for a sequence of forecast horizons $t_1,\ldots,t_T$ in the future ($t_1>t'_k$). 
For forecasting, not only past aggregate load values are available but also additional factors $\vec{x}_{t'_1},\ldots,\vec{x}_{t'_k}$, where the vector  $\vec{x}_{t'}=(x_{t',1},\ldots,x_{t',p})$ summarizes $p$ different explanatory variables. Typical information that is summarized in $\vec{x}_{t'}$ is for example the  temperature, the hour of the day or the season (cf. Section \ref{sec:ForecastMethods}).

\subsubsection{Direct Forecasting} \label{sec:ForecastDirect}

Two variants are distinguished. The first variant is called \textit{direct} forecasting that predicts each $\Lallt$ of the prediction period based on the past regional aggregate values $\mathcal{L}_{t'_1},\ldots,\mathcal{L}_{t'_k}$ and other factors $\vec{x}_{t'_1},\ldots,\vec{x}_{t'_k}$. Note that no zonal aggregate loads are available for forecasting. The corresponding forecast value is denoted by $F_{\mathrm{direct,t}}$.

\subsubsection{Hierarchical Forecasting}\label{sec:ForecastHierarch}

The second variant is called \textit{hierarchical} forecasting that is allowed to use the zonal aggregate loads. First, the aggregate load per zone is predicted and the prediction of the region is obtained by the sum of the predicted zonal aggregates. For each zone $z$, based on the past values $L_{z,t'_1},\ldots,L_{z,t'_k}$ and other factors $\vec{x}_{t'_1},\ldots,\vec{x}_{t'_k}$, future values $L_{z,t_1},\ldots,L_{z,t_T}$ are forecasted. With the forecasts of the zonal aggregates denoted as $F_{z,t_1},\ldots,F_{z,t_T}$, the overall aggregate $\Lall$ is then estimated
\begin{equation} 
\label{equ:overall_aggragate}
\mathcal{F}_{t}=\sum \limits _{z=1}^Z F_{z,t}\quad .
\end{equation}
Hierarchical forecasting is assumed to be beneficial when the loads of different zones must be predicted differently depending on the forecast inputs. For example, loads exhibit more or less distinctive maximum values also at different times of the day. The temperature might as well be described more accurately for smaller, more homogeneous zones.

Hierarchical forecasting with differentially private, perturbed data is effectively the same problem as before with the difference that for each zone instead of the exact aggregates $L_{z,t'_1},\ldots,L_{z,t'_k}$ solely perturbed aggregates $\hat{L}_{z,t'_1},\ldots,\hat{L}_{z,t'_k}$ are used. Since resulting forecast of the regional aggregate then depends on the amount of added noise, it is denoted as $F^\lambda$.

\subsection{Evaluation of Forecasting}
\label{subsec:evaluation}

As stated before we need to add noise to the original data to achieve Differential Privacy. The noise added to the aggregate of each zone yields to the Differential Privacy property of the aggregate load data of each zone $z$. Due to the immunity against post-processing (cf. Section~\ref{subsec:differential-privacy}) the overall aggregate load that is computed as the sum is differentially private.

It is clear that in a practical setting, perturbed data can only be of use if the noise does not destroy the prediction performance of the data. For calculating the utility we compare the direct, non-hierarchical forecast with the hierarchical forecast using differentially private load data of the zones.

The utility of the forecasts for loads of the period $t_1,\ldots,t_T$ is assessed by two error measures. Firstly, the commonly used mean absolute percent error $MAPE$, a scale-free error measure that enables a comparison of our results with results for other datasets.
Secondly, the mean absolute error $MAE$ that allows to compare  different forecasting methods for the same (GEFCom) data.
Both error measures are computed according to their names, i.e.,
\begin{eqnarray}
MAE &:=& \frac{1}{T} \sum_{t=t_{1}}^{t_{T}} \left| F_{t}-L_{t} \right|\quad, \label{eqn:mae}\\
MAPE &:=& \frac{1}{T} \sum_{t=t_{1}}^{t_{T}} \left| \frac{F_{t}-L_{t}}{L_{t}} \right|\quad.
\end{eqnarray}
This way the error can be assessed both for forecasting the aggregate load $L_z$ of a zone $z$ and the overall aggregate load $\Lall$.

We define utility $u^{\lambda}$ as the relative gain we achieve by switching from non-hierarchical to hierarchical forecast with perturbed data,
\begin{equation} \label{eq:utility}
u^{\lambda} := \frac{MAE_\mathrm{direct}-MAE^{\lambda}}{MAE_\mathrm{direct}}\quad .
\end{equation}
The error measure $MAE^{\lambda}$ uses the forecasts $F^{\lambda}$ of the hierarchical forecast with perturbed data for the overall aggregate load $\Lall$, where $MAE_\mathrm{direct}$ is the error of the direct load forecast for the overall aggregate load $\Lall$. Since the regional aggregate values are known in any case, direct forecasting can always be done. Therefore, hierarchical forecasting only makes sense when it is better than direct forecasting. Consequently, when the perturbation factor $\lambda$ becomes too large it causes the error $MAE^\lambda$ to exceed the direct forecasting error $MAE_\mathrm{direct}$ and the utility becomes negative $u^{\lambda} \leq 0$.

\subsection{Differentially Private Metering Process}
\label{sec:differentially-private-metering-process}

In this work, we strive to bring energy providers in the position to train load forecast models on differentially private aggregated data from electricity customers. Differential Privacy is required due to possible insufficiencies of pure aggregation for privacy protection \cite{Dwork06b, Buescher17a}. 

As stated previously, energy providers desire to limit deviation of their forecasting algorithms due to differentially private noise added to load forecasting training data by specifying an upper bound for acceptable forecasting error. In turn, this will lead to an upper bound for acceptable noise scales $\lambda$ which is needed according to the Laplace mechanism for achieving Differential Privacy. 
More specifically, by Theorem~\ref{def:lap-mech} each household is provided $\frac{\Delta f}{\lambda}=\epsilon$-Differential Privacy. 
The application of the Laplace mechanism results in three challenges in the scenario of this paper. 

Firstly, we do not want to have perturbation $L_{z,t'_1},\ldots, L_{z,t'_k} \to \hat{L}_{z,t'_1},\ldots, \hat{L}_{z,t'_k}$ done by the energy provider to avoid assumptions about trustworthiness.
Instead, we desire perturbation to be performed at the data sources directly, i. e. a smart meter adds noise itself for each point in time $t'$. Following Lemma~\ref{def:gamma}, we realize this by decomposing the Laplace noise into the gamma noise for distributed noise generation at household level as stated in equation~\eqref{equ:noisy-household}. The provider has to compute the sum for each zone to obtain the noisy total consumption, see equation~\eqref{equ:noisy-sum}.

\begin{align}
\hat{l}_{z,i,t'} &= l_{z,i,t'} + (\mathcal{G}_1(n_z, \lambda) - \mathcal{G}_2(n_z, \lambda))\label{equ:noisy-household}\\
\hat{L}_{z, t'} &= \sum_{i=1}^{n_z} \hat{l}_{z,i,t'} = L_{z, t'} + \mathcal{Z}(\lambda) \label{equ:noisy-sum}
\end{align}

Secondly, the training data is represented by a time series $t'_1,\ldots, t'_k$ of each electricity customer's energy data, i.e., involving always the same set of households.
Consequently, privacy decays over time as more information is revealed. For measuring the accumulated privacy loss, we apply Theorem~\ref{def:adv-composition} to obtain the total privacy loss $\tilde{\epsilon}_{\tilde{\delta}}$ as a function of $\epsilon,\delta$ and time $k$.

Thirdly, the accumulated privacy guarantee, $\tilde{\epsilon}_{\tilde{\delta}}$, is hard to interpret for consumers (i.e., electricity customers). Our envisioned process addresses this by translating $\tilde{\epsilon}_{\tilde{\delta}}$ into an interpretable risk $\rho$ by equation~\eqref{equ:eps-rho}. $\rho$ represents the upper bound for the confidence of an adversary trying to detect the presence of a single household in $\hat{L}_{t',z}$. We have almost perfect privacy if an attacker is unable to confidently distinguish whether a household contributed to the sum or not, i.e., $\rho \approx 0.5$ (random guessing). In contrast, if $\rho \approx 1$ the privacy level is extremely low. 
To provide a reasonably good protection, we aim to bound the confidence at $\rho = 0.6$, meaning that even in worst case situations an adversary is not able to identify that a household contributed with more than 60\% confidence.

Our process of applying Differential Privacy has several benefits. The energy provider does not have to perform any perturbation as noise is added locally by each meter and adds up to noise following the Laplace mechanism. In addition, providers can select the amount of noise $\lambda$ they tolerate with regard to their forecasting algorithms. $\lambda$ then gets propagated to households who resolve it to their corresponding $\rho$ to see how much data privacy the energy provider actually ensures.


\section{Experiments and Results}
\label{sec:experiments}

In this section three different models for forecasting the GEFCom data set (Section~\ref{sec:GefCom}) are trained. 
After confirming the correctness of the implementations by applying the forecasts to unperturbed data, the sensitivity of the forecasting performance on the Laplacian noise of different scales $\lambda$ is assessed. Because the noise scale $\lambda$ does not lend itself to describe the achieved privacy in a comprehensive way, such a description is developed in Section~\ref{subsec:application-of-differential-private-mechanism} based on the Differential Identifiability notion. Using all of the above, the privacy-utility trade-off will be described.

\subsection{Forecast results}

We re-implemented Hong's Linear Regression Benchmark Model and CountingLab's forecast model. For sake of simplicity we omitted 2 of the improvements of CountingLab's model. Lloyd's method did not need to be implemented because he provided the source code \footnote{Source code available at \texttt{https://github.com/jamesrobertlloyd/GEFCOM2012}}. Only adaptions facilitating the handling of many different input files have been necessary. 

\begin{figure}[p]
	\centering
	\begin{subfigure}[b]{1.0\textwidth}
		\centering
		\includegraphics[width=\figurewidth]{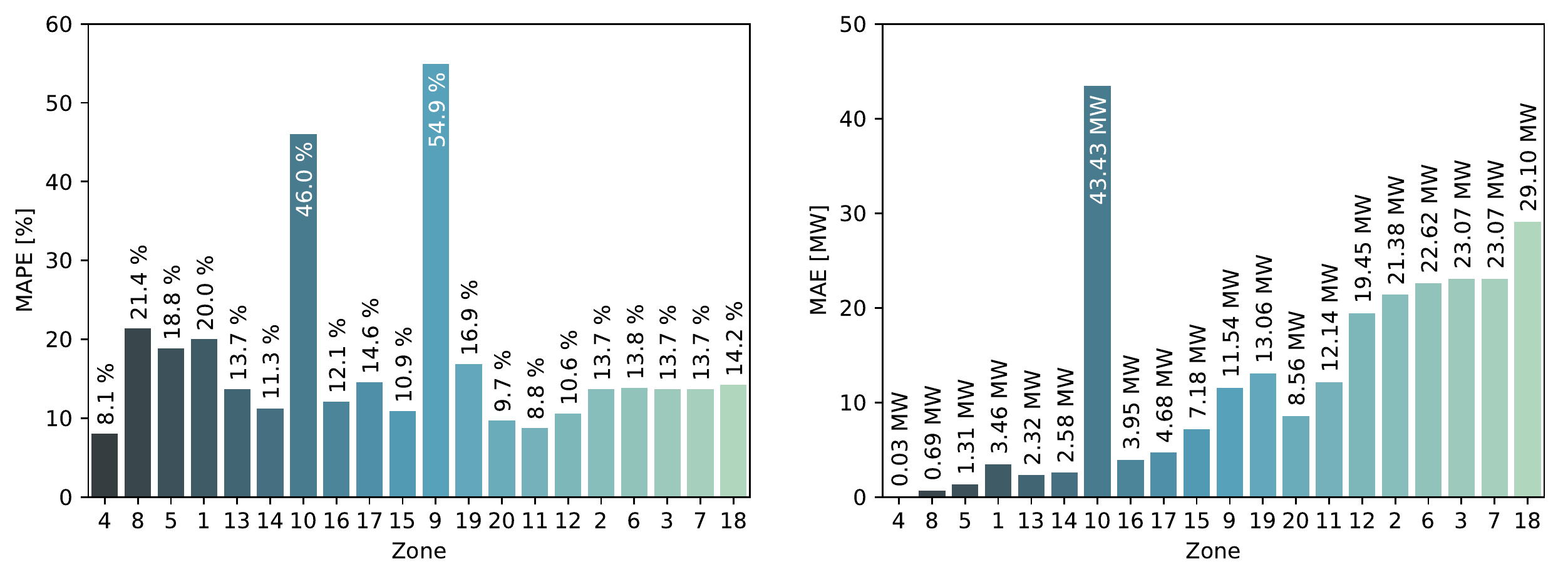}
		\setlength{\abovecaptionskip}{0pt plus 5pt minus 2pt}
		\caption{Hong's Benchmark Model with unperturbed data.}
		\label{fig:mlr_no_noise_results}
	\end{subfigure}
	\begin{subfigure}[b]{1.0\textwidth}
		\centering
		\includegraphics[width=\figurewidth]{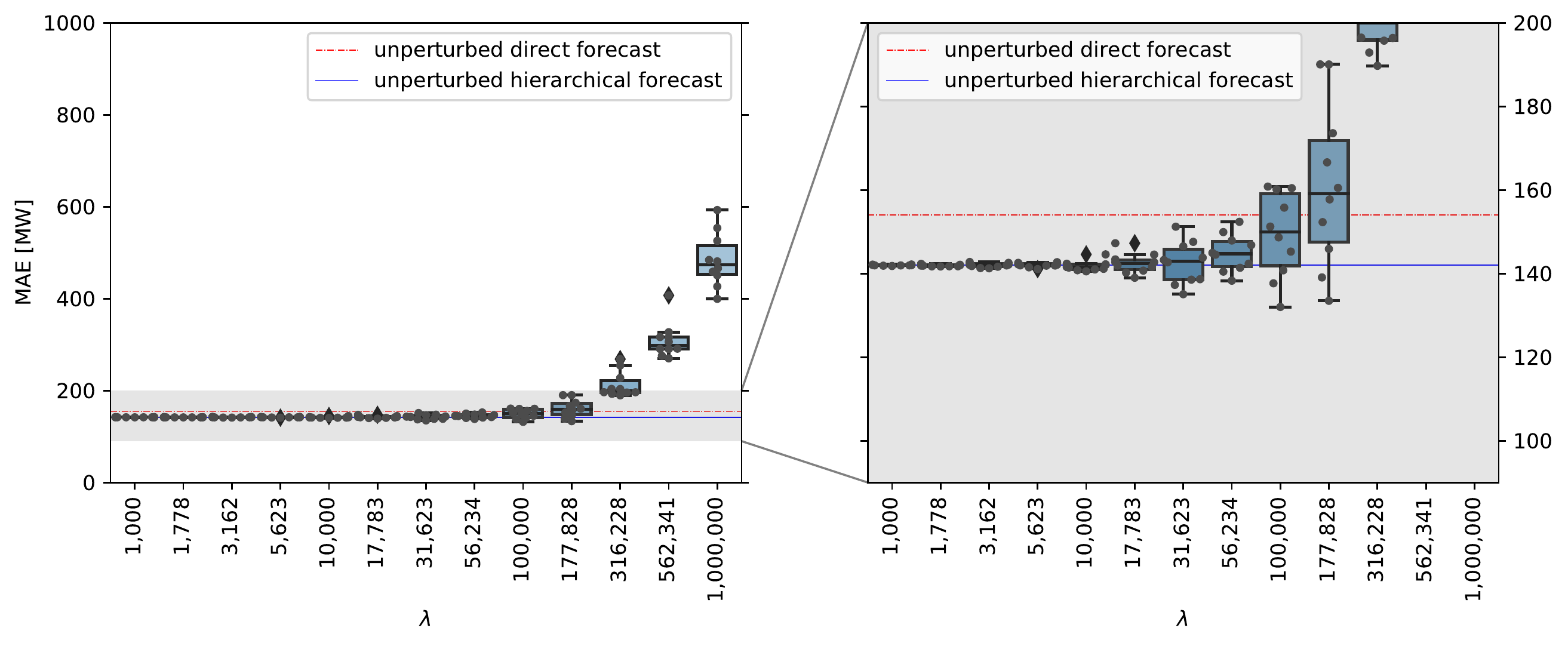}
		\setlength{\abovecaptionskip}{-2pt plus 5pt minus 2pt}
		\caption{Hong's Benchmark Model with perturbed data.}
		\label{fig:mlr_maes_10runs}
	\end{subfigure}
	\begin{subfigure}[b]{1.0\textwidth}
		\centering
		\includegraphics[width=\figurewidth]{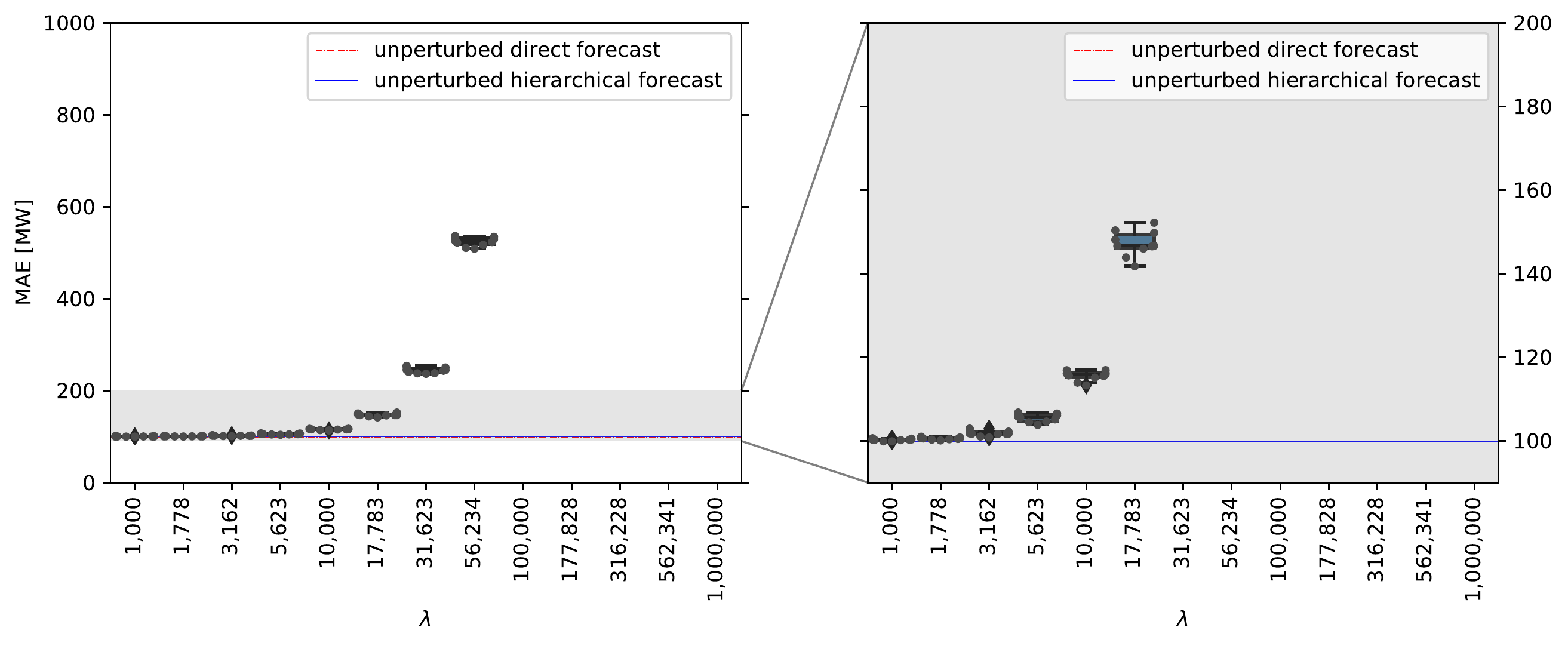}
		\setlength{\abovecaptionskip}{-2pt plus 5pt minus 2pt}
		\caption{CountingLab's Model with perturbed data.}
		\label{fig:clabMaesAggregatePerturbed}
	\end{subfigure}
	\begin{subfigure}[b]{1.0\textwidth}
		\centering
		\includegraphics[width=\figurewidth]{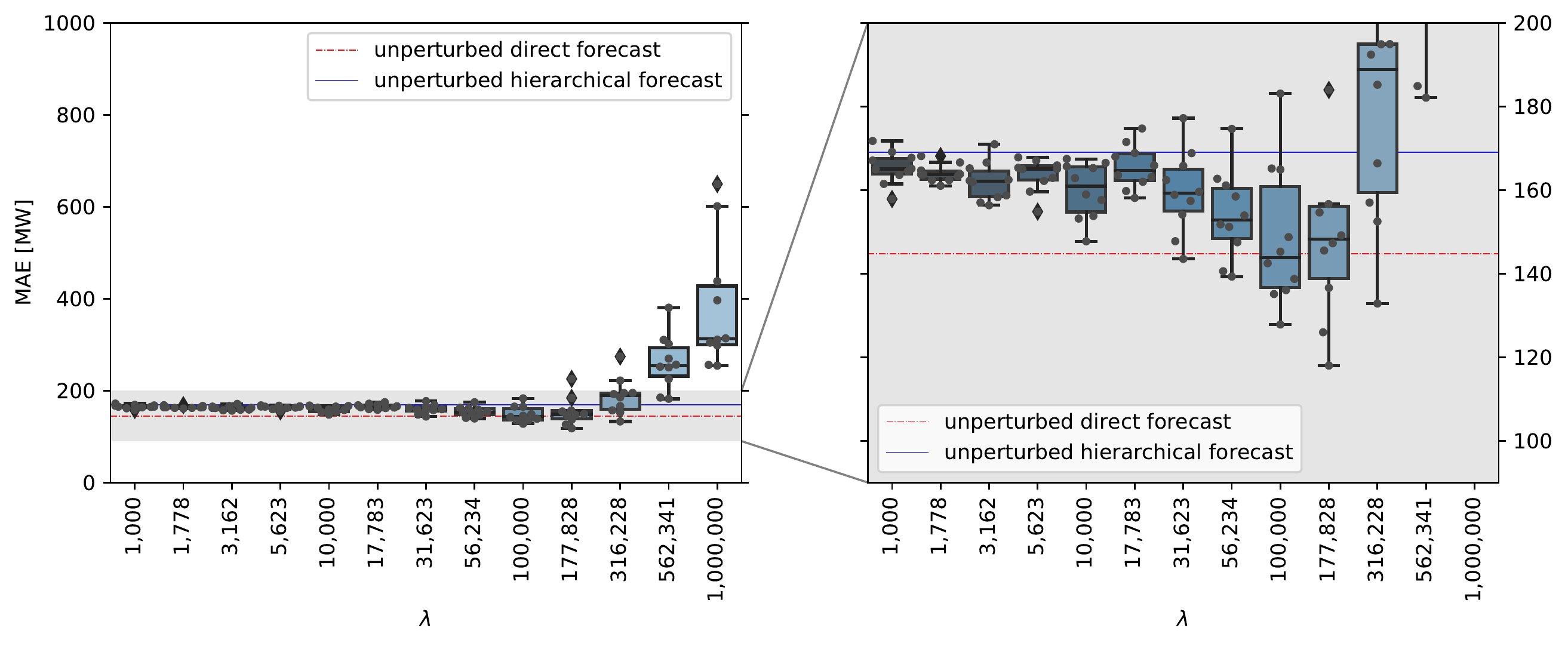}
		\setlength{\abovecaptionskip}{-2pt plus 5pt minus 2pt}
		\caption{Lloyd's Model with perturbed data.}
		\label{fig:LloydMaesAggregate}
	\end{subfigure}
	\caption{(a): MAPE (left) and MAE (right) of the benchmark model for different zones (ordered by average load). Panels (b)-(d): Forecasting errors of the forecasting models with increasing levels of noise. Blue, solid lines: MAE of the hierarchical models, each without noise. Red, dashed lines: MAE of the direct forecast models.}
\end{figure}

Firstly, we verified the correctness of the implementation for unperturbed data. 
The MAPE and MAE of the non-perturbed forecast by Hong's model for each zone are depicted in Figure~\ref{fig:mlr_no_noise_results}.
The zones are sorted by their average load from left to right.
Zone 9 and 10 have prominently high errors.
As Figure~\ref{fig:zonestats} shows, the outliers in Zone 9 indicate metering errors or power outages.
In Zone 10, the average monthly consumption suddenly tripled starting January 2008, indicating a change of the grid configuration.
As the forecasting time period is after January 2008, this may be the cause of the high forecasting errors.

Similarly, both CountingLab's and Lloyd's forecast models are bad for zones 9 and 10. However, for both of these models the errors are on average smaller for the remaining zones than for the benchmark model.

A comparison between unperturbed direct forecast and the unperturbed hierarchical forecast shows that the hierarchical forecast for the benchmark method lowers the average error by 12 MW. This results in a utility of 7.8\% (first line of Table~\ref{tab:privacy-levels-adaptive composition}) and means that our privacy mechanism should not introduce additional errors much above 12 MW in order to avoid too negative utilities. Surprisingly, the hierarchical forecast is worse than the direct forecast for the other two models resulting in negative utilities.

Now, the impact of varying levels of noise on the forecast performance is evaluated.
Figure~\ref{fig:mlr_maes_10runs} shows the forecasting error of perturbed hierarchical forecasting of Hong's Benchmark Model using increasing levels of perturbation.
We trained the models and ran the forecast 10 times each with different random seeds.
In some cases, the error even decreases.
The red line indicates our utility-limit of 12 MW above unperturbed error (blue line).
With $\lambda = 56,234$, all runs still stayed below this limit.
Starting at $\lambda = 100,000$, some runs start to show higher error than the unperturbed direct forecast.

While the performance of CountingLab's forecast is better than the benchmark model for unperturbed data, the performance is highly negatively affected by the noise. This can be seen in Figure~\ref{fig:clabMaesAggregatePerturbed} where the MAE quickly rises with $\lambda$.

The main difference between CountingLab's method and the benchmark model lies in the construction of many small models that use a smaller amount of data, each. It seems plausible that noise can have a greater negative effect on such approaches.

The MAPE and MAE of Lloyd's forecast with perturbed data for each zone are depicted in Figure~\ref{fig:LloydMaesAggregate}. Surprisingly, the forecast first improves for some amount of noise, reaches a minimum at $\lambda=177,828$ and then rises quickly. 

This behavior can be attributed to the gradient boosting model which also has the highest weight (0.765) in the ensemble averaging process (not shown). Since the inputs of the gradient boosting model do not include any load values (compare Section~\ref{sec:modelLloyd}), Differential Privacy acts as output noise which has been shown to potentially improve a model by Breiman et al. \cite{Breiman00a}. As the benchmark model did, the third classifier of the ensemble, the Gaussian Process model, degrades monotonically and finally rather quickly with increasing $\lambda$ (not shown).
The bad reaction upon noise of the Gaussian Process is plausible since the model heavily relies on a limited amount of 500 load values which corresponds to three weeks of data.
However, since it only has a weight of 0.135 the behavior of the gradient boosting model dominates for small $\lambda$. 

\subsection{Application of Differential Privacy}
\label{subsec:application-of-differential-private-mechanism}

While we conceptually presented the integration of Differential Privacy into smart metering load forecasting in Section~\ref{sec:differentially-private-metering-process}, we provide an evaluation of the implementation in the following. 

As initial step, we let an energy provider set utility bounds by choosing the noise scale $\lambda$ in dependence of the acceptable loss in utility, i.e., forecast accuracy.
In the next step we fix $\Delta f = 48$ kW as \textit{global} $\Delta f$  (i.e., maximum power consumption), which is the maximum power limit of 3-phased circuits in German residential homes.
Based on $\lambda$ and $\Delta f$, a global privacy guarantee of ($\epsilon, 0$)-Differential Privacy \eqref{eq:DP} is provided by each individual load aggregate $\hat{L}_{z,t}$ using the Laplace mechanism \eqref{eq:Laplace}.

However, this theoretical restriction is far from being reached in practice. Thus, households may exchange the global $\Delta f$ by a smaller, \textit{local} $\Delta f$ to identify their actual privacy guarantee. 
Considering the same $\lambda$, since $\epsilon=\Delta f/\lambda$, households may actually enjoy a stronger (smaller $\epsilon$) protection against presence detection under their local $\Delta f$.
However, the $\epsilon$ guarantee then only applies to loads within the local interval and does not keep an attacker from finding out about the bounds of that local interval.
In the end, it is a matter of interpretation whether one relies on a very theoretical protection guarantee or a more realistic relaxation. To illustrate the impact, we vary $\Delta f$\footnote{Local $\Delta f$ are based on statistics retrieved from the CER data set \cite{CER12} as the GEFCom data set only contains aggregates. See Appendix~\ref{app:local-sensitivities} for details.} according to Table~\ref{tab:selected-sensitivity} for our scenario.

\begin{table}
	\centering
		\caption{Selected $\Delta f$ and according reasoning based on \cite{CER12}}
	\begin{tabularx}{\columnwidth}{cX} \toprule[1.5pt]
		$\Delta f$ & Argument \\
		\midrule[0.25ex]
		 $7.57$ kW & $90$th percentile of highest power demands recorded per hour\\
		 $10.05$ kW& $99$th percentile of highest power demands recorded per hour\\
		 $15.36$ kW& highest power demand per hour in the whole data set \\
		 $48.00$ kW& maximum power demand fused in German residential homes\\
		\bottomrule[1.5pt]
	\end{tabularx}
	\label{tab:selected-sensitivity}
\end{table}

When continuously releasing information by computing $\hat{L}_{z,t'_1},\ldots,\hat{L}_{z,t'_k} $ a composition theorem has to be applied as each $\hat{L}_{z,t'}$ relates to the same set of individuals (i.e., households). The GEFCom data set consists of $k=38,070$ hourly load recordings, thus we have almost $40,000$ composition steps. 
For large k, however, k-fold adaptive composition (Section~\ref{subsec:differential-privacy}) is a tight estimation of the privacy loss. By fixing some very small $\hat{\delta}$, the growth of a composed $\tilde{\epsilon}_{\tilde{\delta}}$ no longer \eqref{eq: epscomposition} depends linearly on $k$. 
We set $\tilde{\delta} \le \frac{1}{|D|}$, where in the worst case w.r.t. the GEFCom data set $|D|$ is the number of all households in the US in 2013\footnote{Estimated $117,716,237$ by the U.S. Census Bureau: \url{https://www.census.gov/quickfacts/fact/table/US/HSD410216}}, i.e., $\tilde{\delta} = \frac{1}{117,716,237} \approx 10^{-9}$.
In the end, each household is protected by ($\tilde{\epsilon}_{\tilde{\delta}}$, $\tilde{\delta}$)-Differential Privacy.

\begin{figure}
	\centering
	\includegraphics[width=0.9\linewidth]{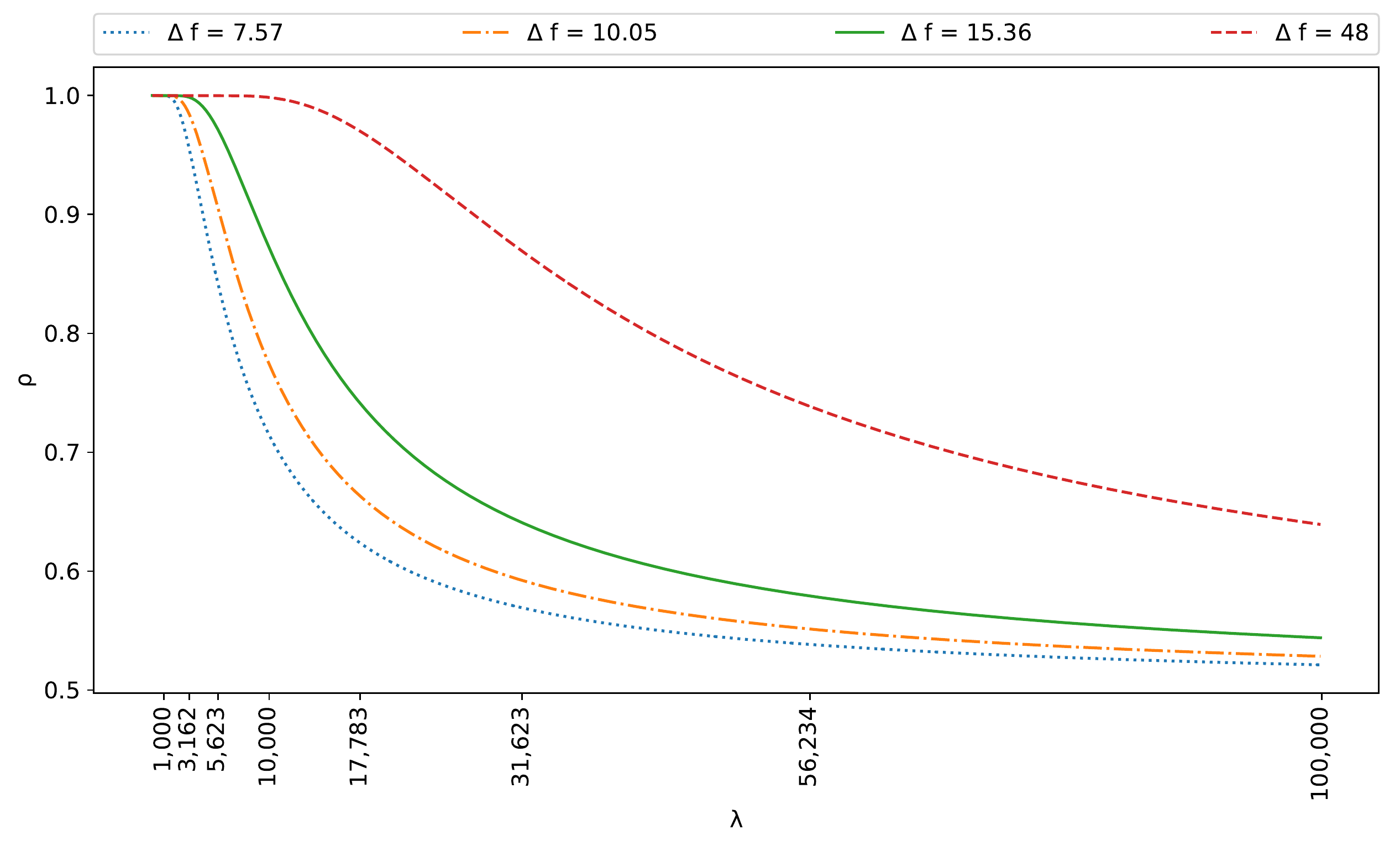}
	\caption{Confidence $\rho$ and forecasting MAE for various $\Delta f$ under composition.}
	\label{fig:rho}
\end{figure}

Regarding our aim to express the privacy guarantee in a comprehensible way, $\tilde{\epsilon}_{\tilde{\delta}}$ is transformed into $\rho$ by~\eqref{equ:eps-rho}.
The impact of $\lambda$ on $\rho$ is displayed in Figure \ref{fig:rho} for various $\Delta f$ to illustrate the significant difference in presence detection likelihood when using theoretical worst case power consumption (i.e., $\Delta f = 48$ kW) or realistic maximum demands (i.e., $\Delta f = 15.36$ kW).
Lowering $\Delta f$ to more realistic values causes $\rho$ to decrease and consequently results in stronger protection against presence detection. Thus, for $\lambda \ge 50,000$, households with realistically estimated maximum loads ($\Delta f$) have already acceptable privacy levels. At $\lambda = 100,000$, even the theoretical worst case of 48 kW approaches the desired $\rho = 0.6$ (cf. Section~\ref{sec:differentially-private-metering-process}). 

\begin{table}
	\centering
	\captionof{table}
	{Noise ($\lambda$) and sensitivities ($\Delta f$) lead to $\epsilon$ and interpretable re-identification confidence ($\rho$) for k-fold adaptive composition ($k=38,070$). The first row states utility ($u$) of the hierarchical, unperturbed forecast ($\lambda=0$) over the direct forecast.}
	\begin{tabularx}{\columnwidth}{*{8}{>{\centering\arraybackslash}X}} \toprule[1.5pt]
		\multirow{2}{*}{$\lambda$} & \multirow{2}{*}{$\Delta f \;\text{[kW]}$} & \multirow{2}{*}{$\epsilon$} & \multirow{2}{*}{$\tilde{\epsilon}_{\tilde{\delta}}$} & \multirow{2}{*}{$\rho$} & \multicolumn{3}{c}{$u^{\lambda}$}\\
				& & & & & CountingLab & Lloyd & Benchmark\\
		\midrule[0.25ex]
		0  & - & - & - & - & -1.53 & -16.81 & 7.80\\\hline
	10,000 &   7.57 &  0.00076 &  0.92 &  0.72 & \multirow{4}{*}{-17.62} & \multirow{4}{*}{-10.48} & \multirow{4}{*}{7.98}\\
	10,000 &  10.05 &  0.00100 &  1.23 &  0.77 &  & & \\
	10,000 &  15.36 &  0.00154 &  1.92 &  0.87 &  & & \\
	10,000 &  48.00 &  0.00480 &  6.46 &  1.00 &  & & \\\hline
	56,234 &   7.57 &  0.00013 &  0.15 &  0.54 & \multirow{4}{*}{\textbf{-433.29}} & \multirow{4}{*}{\textbf{-6.49}} & \multirow{4}{*}{\textbf{5.94}}\\
	56,234 &  10.05 &  0.00018 &  0.21 &  0.55 &  & & \\
	\textbf{56,234} &  \textbf{15.36} &  \textbf{0.00027} &  \textbf{0.32} &  \textbf{0.58} &  & & \\
	56,234 &  48.00 &  0.00085 &  1.04 &  0.74 &  & & \\\hline
	100,000 &   7.57 &  0.00008 &  0.08 &  0.52 & \multirow{4}{*}{-1084.80} & \multirow{4}{*}{-2.76} & \multirow{4}{*}{3.10}\\
	100,000 &  10.05 &  0.00010 &  0.11 &  0.53 &  & & \\
	100,000 &  15.36 &  0.00015 &  0.18 &  0.54 &  & & \\
	100,000 &  48.00 &  0.00048 &  0.57 &  0.64 &  & & \\
		\bottomrule[1.5pt]
	\end{tabularx}
		\label{tab:privacy-levels-adaptive composition}
\end{table}

The trade-off between privacy and utility is shown in Table~\ref{tab:privacy-levels-adaptive composition}.
Both CountingLab's and Lloyds's model work better for the direct than for the hierarchical setting. In contrast, the hierarchical benchmark forecast outperforms its direct counterpart. Thus, only the benchmark model is a suitable candidate for differential privacy. This is an interesting and unexpected result (note that although the performance of Lloyd's forecast improves with limited amount of noise it never has a positive utility). The desired presence detection confidence region $\rho \leq 0.6$ is achieved for the benchmark model for $\lambda=56,234$ with $\Delta f = 15.35$ and  offers a positive utility of 5.94\% with respect to the direct forecast. Thus, a setting has been found where both, privacy and utility, have been reached. 
The authors want to highlight that they assume communication of individual, understandable presence detection risk $\rho$ based on individual $\Delta f$ as crucial to foster consumer acceptance of privacy-preserving techniques.

\section{Related Work}
\label{sec:related}

One of the first works to discuss and demonstrate privacy issues with smart metering was from Molina-Markham et al. \cite{Molina10}.
Later, Greveler et al. \cite{Greveler12} demonstrated that the TV program can be inferred based on high-resolution load monitoring.
Most recently, Rafsanjani et al. \cite{Rafsanjani18} showed empirically that the occupancy of a commercial building can be estimated based on high-resolution energy consumption data with an accuracy above 95\%.

Two prominent use-cases of smart metering data are electricity consumption billing and real-time monitoring for grid operations.
For billing exact fees are important, so due to the addition of noise Differential Privacy has only rarely been applied \cite{Danezis11a}. Typically, privacy is improved by disclosing only the necessary information for the business process, which is, at best, the final cost of each individual.
Molina-Markham et al. \cite{Molina10}, Rial and Danezis \cite{Rial2011_privacy-preserving}, and Jawurek et al. \cite{Jawurek2011_plug-privacy} use Zero-Knowledge Protocols to provide a privacy-preserving billing.

For real-time electricity monitoring, information aggregated over a geographical or topological grid area is sufficient.
The privacy enhancing approaches for this use-case are mostly based on mixing networks which are partially backed by homomorphic encryption.
Examples include works by Li et al. \cite{Li10a}, Garcia and Jacobs \cite{Garcia2011_privacy-friendly}, Defend and Kursawe \cite{Defend2013_implementation}, and Finster and Baumgart \cite{Finster2014_smarter}.

All the approaches so far require the metering infrastructure to be designed in a specific way.
As a privacy self-defence mechanism, a grid customer could resort to load obfuscation.
Load Obfuscation physically manipulates the load profiles of households by using battery storage systems or controllable loads and generators.
Examples are Kalogridis et al. \cite{Kalogridis2010_privacy} and McLaughlin et al. \cite{Mclaughlin2011_protecting}, who leverage batteries to shift loads,
Chen et al. \cite{Chen2014_combined}, who controls Combined Heat and Power plants,
and Egarter et al. \cite{Egarter2014_load}, who uses appliances in an energy management approach to protect privacy.

The closest related to our work are differentially private smart metering concepts.
Ács and Castelluccia were the first to apply Differential Privacy on smart metering data.
In their work \cite{Acs11a}, the Laplace mechanism is applied distributedly using Gamma distributions before the data is mixed with other Smart Meters in an aggregation group.
Bao and Lu \cite{Bao15a} investigated further the security and fault tolerance properties of the aggregation and mixing protocol.
Eibl and Engel \cite{Eibl16a} introduced post-processing to be applied on the perturbed data to improve the utility while still guaranteeing the same privacy level.
They also discuss the required number of households in an aggregation group in order to be useful to the data analyst.
Barbosa et al. \cite{Barbosa16} also discussed filtering techniques to improve utility after the noise has been added to the aggregate.
Their work evaluates the protection of individual appliances in single households by considering multiple device sensitivities in load profiles and by using Differential Identifiability. However, they do not address the compatibility condition $m=2$ to allow utilizing Differential Identifiability in Differential Privacy scenarios.
Besides Differential Identifiability, another method for rationally choosing $\epsilon$ was proposed in \cite{Hsu14}. 
Yet, this approach is purely economically driven and introduces a handful of new parameters depending again on subjective assumptions on a given scenario. 
As our focus is primarily on security, we decide to further analyze Differential Identifiability and its belief model only.
From Ács et al. \cite{Acs11a} we borrowed the idea to generate noise scaled to the Laplace mechanism with the gamma distribution. We carried their work further by connecting it to Differential Identifiability and load forecasting with utility guarantees.

\section{Conclusion and Outlook}
\label{sec:Conclusion}

In this paper, we discussed that energy providers are interested in smart metering data to refine the forecast of domestic loads of their customers.
As this conflicts with the privacy loss incurred by the acquisition of individual load profiles, we designed a differentially private metering process based on building blocks already proposed in previous works.
Using three well-documented load forecasting approaches, we evaluate whether using smart metering data provides an actual benefit for the energy provider.
We found out that this is not always the case and that the forecasting approaches are variously susceptible to noise.
If smart metering data actually provides a utility to the energy provider, Differential Privacy allows to gradually trade-off utility against forecasting performance.
Our results show that for one forecasting approach, reasonable utility can be reached while providing a strong privacy guarantee.
In that case, Differential Identifiability even provides an intuitive interpretation of the amount of privacy loss.

Several important points have to considered when our concept is to be applied safely in practice:
Firstly, there is no privacy guarantee for individual smart metering data of a \textit{single} household.
In particular, the sum of individual load and Gamma noise is still sensitive, therefore secure aggregation with other households is crucial.
That is why we stated homomorphic encryption and masking or mixing as minimum requirement (cf. Section \ref{sec:problem}).
Secondly, we considered privacy guarantees from a static snapshot of the scenario when the energy provider has collected approximately 4.5 years of zonal load profiles.
Applying our approach in practice continuously would mean that the privacy guarantee is stronger if less than 4.5 years of data was collected from you.
After 4.5 years our evaluated privacy guarantees would slowly degenerate.
Thirdly and tightly connected to the second point, the historic and forecasted load profiles of our used data set were given with hourly read-out 
intervals.
However in Europe, load profiles are acquired in 15 minute intervals.
Our findings also apply to this case with the only difference that the privacy guarantee would hold for slightly more than one year instead of 4.5 years.
Finally, if the privacy level offered by the energy provider is not high enough to protect the electricity usage of the whole household, the protection can still be interpreted for single household appliances.
In this case, one has to be aware that the usage of this single appliance is not allowed to correlate to the (parallel) usage of other appliances.

There are several natural extensions to the presented work:
Firstly, for utility evaluation, we used three well-documented point forecasting methods.
Point forecast outputs only a single most-likely load value for one time interval.
An extension to this work would be to evaluate differentially private metering with probabilistic forecasting methods (cf. \cite{Hong16}).
Secondly, our concept pertubates and transmits the complete zonal time series to the energy provider and the forecasting model training is performed by the energy provider.
In the future, we plan to integrate Differential Privacy directly into a distributed model training approach on the customer side using \textit{objective-function perturbation} for less privacy loss and tighter guarantees.
Thirdly, lowering the local sensitivity by minimizing the household's peak load leads to a stronger privacy level.
Incidentally, automatic energy management systems like the ones described in \cite{Egarter2014_load} and \cite{Mauser16} are able to shift controllable loads or control battery storages and combined heat and power plants to facilitate this idea.
Fourthly, with the continual release of load data in practice, the privacy loss quantified by $\epsilon$ would slowly add up over the course of time.
To be aware of one owns privacy situation, one needs to keep track of how much privacy was already leaked to which party.
The data custodian proposed in~\cite{Rigoll2017acfauedms} provides such an accounting service.
Finally, the perturbed data could be filtered (e.g., using moving average or Kalman filters) to compensate the noise as already proposed in \cite{Bao15a,Eibl16a}.

As final remark, using local sensitivities creates the incentive to limit one own's energy consumption due to privacy protection interests.
Although this behavior may be beneficial to the electric grid, this would not be in the spirit of informational self-determination.
That is why using the global sensitivity instead of local sensitivities should be preferred.

\subsubsection*{Acknowledgments}

This work has received funding from the European Union's
Horizon 2020 Research and Innovation Programme under grant agreement No. 653497 PANORAMIX
and from the Federal Ministry of Education and Research (BMBF) under funding No. 16KIS0521 KASTEL-SVI.

\small
\bibliographystyle{bmc-mathphys} 
\bibliography{../00-Bib/final}      

\appendix
\section{Local Sensitivities of Households}
\label{app:local-sensitivities}
For the Adaptive Composition, one has to regard the peak load (power) within a read-out interval. There are 
 \textit{technical} bounds on the instantaneous electric power usage we can use to derive global sensitivities for our privacy model.
In Germany, a household is usually protected with a 63 A contractor and the household is connected to all three AC-phases (cf. Section 15.2 in \cite{Kasikci2013} and DIN 18015 Part 1).
With a nominal voltage of 230 V and an acceptable over-voltage of 10\%, we get that the highest electrical power consumption a German household could have is $ P_{peak} = 230V \cdot 1.10 \cdot 63A \cdot 3 \approx 48$ kW.
The global sensitivity $\Delta f = 48$ kW for a read-out interval is very likely to be higher than actual peaks households' power demand. Consequently, we calculate individual households' risk with their corresponding local sensitivities in the privacy evaluation. This enables households to derive their individual Differential Privacy guarantee. To obtain these values, we unfortunately cannot use the GEFCom data set due to missing information on load recordings of single households. Thus, we calculated the 90th, 99th and 100th percentile of the highest consumption peaks over all households from the comparable CER electric data set \cite{CER12} to obtain good approximation of a realistic maximum.

\end{document}